\documentclass[12pt,a4paper]{article}

\usepackage[T1,T2A]{fontenc}
\usepackage[utf8,utf8x]{inputenc}

\pdfoutput=1
\newcommand{\be}{\begin{equation}}
\newcommand{\ee}{\end{equation}}
\newcommand{\bea}{\begin{eqnarray}}
\newcommand{\eea}{\end{eqnarray}}
\newcommand{\beas}{\begin{eqnarray*}}
\newcommand{\eeas}{\end{eqnarray*}}
\newcommand{\ba}{\begin{array}}
\newcommand{\ea}{\end{array}}
\newcommand{\nn}{\nonumber}

\newcommand{\bt}{\begin{table}}
\newcommand{\ve}{\varepsilon}

\newcommand{\vart}{\vartheta}

\newcommand{\al}{\alpha}
\newcommand{\ga}{\gamma}

\newcommand{\Ga}{\Gamma}

\newcommand{\ka}{\kappa}
\newcommand{\la}{\lambda}
\newcommand{\La}{\Lambda}
\newcommand{\na}{\nabla}

\newcommand{\si}{\sigma}

\begin{document}

\title{ {
\bf  Affine-Goldstone/quartet-metric gravity and beyond
}}
\author{Yury~F.~Pirogov
\\
\small{Theory Division, 
Institute for High Energy Physics of
NRC Kurchatov Institute,}
\small { Protvino,
Russia}
\\
}
\date{}
\maketitle

\begin{abstract}
\noindent 
As a group-theoretic foundation of gravity, 
it is  considered an affine-Goldstone      
nonlinear model based upon the nonlinear  
realization of the global affine symmetry 
spontaneously  broken  at the Planck scale to the Poincare symmetry. 
It is shown that  below this  scale the model  justifies  and elaborates   
an  earlier  introduced effective field theory 
of  the  quartet-metric gravity incorporating 
the  gravitational  dark  substances emerging
in addition to the tensor graviton. 
The prospects for subsequent going beyond the  nonlinear model  above   
the Planck scale are indicated.\\

\noindent
{\bf PACS:} 04.50.Kd Modified theories of gravity,   95.35.+d Dark matter, 
{95.36.+x} Dark energy. 

\end{abstract}

\section{Introduction: GR and beyond}

General Relativity (GR) is the well-stated contemporary basis of gravity 
remaining up-to-date   in a position to successfully cope with 
the bulk of the  astrophysical and 
cosmological manifestations of gravity.
Nevertheless,  it may be argued that 
an underlying nature of gravity beyond GR is  still  obscure.
The reason may come from the impressive recent achievements 
in the observational astrophysics and cosmology.
Namely, the   advent 
in the Universe of the elusive dark substances, 
such as dark matter (DM)  and dark energy (DE), 
in an  amount  disproportionately large 
($\sim95\%$  of the total energy) compared to the 
ordinary matter ($4\div5 \%$),\footnote{See, e.g.,~\cite{PDG}.}  
with   their  nature remaining moreover  completely obscured, causes 
some (still  mainly theoretical) tension within GR.
The predominant abundance of such 
the {\em ad hock}  dark substances, though  
quite legitimate  in the GR framework,  may  be a hint from the Nature  at 
a necessity of going beyond GR,  with the elusive dark substances 
being  nothing but an integral   part of the modified  gravity itself. 

With such an   aim in mind,   as a modification of 
the (metric)  GR
 there was recently proposed in refs.~\cite{Pir1,Pir2}
 the  effective field theory  (EFT) 
of the spontaneously broken quartet-modified GR (or, in short, 
the  quartet-metric  gravity).  
The latter is based originally upon the three physical concepts. 
First, there exist in spacetime the distinct dynamical  coordinates  given 
by a scalar quartet (in the four spacetime dimensions).
Such  coordinates  are ultimately associated with the vacuum treated 
as the dynamical one
{\em on par} with the other fields.
Second, the diffeomorphism symmetry of the quartet-modified 
GR undergoes the spontaneous symmetry breaking (SSB), with 
the scalar quartet playing the role of the Higgs-like 
fields for gravity.
Third, the physical  gravity components arising 
from the gauge ones  of metric due to SSB 
serve as the gravitational dark substance  of the Universe. 
It was  argued that  such a  constructed EFT of gravity may give rise 
to a large variety of manifestations beyond GR,
to consistently study which presents  a  challenge. 
At that, the latter is hampered by the fact,  that by the very construction 
the conventional EFT frameworks  
inevitably contain  a number   of the 
{\em ad hock} assumptions and parameters 
implying, conceivably, a necessity of further  elaborating such a theory,
as well as its  foundations.
 
In the latter  respect note, that GR (like its direct siblings) is  
well-known to be based upon 
the pseudo-Riemannian/metric paradigm of gravity, 
which is conventionally assumed to be an attribute of gravity. 
On the other hand, already long ago there was put forward~\cite{Salam1, Ogi}
an alternative,  group-theoretic/Nambu-Goldstone (NG)  paradigm,  
wherein gravity has  an NG origin  being  
based upon the spontaneous breaking  
at the Planck scale  of the global affine symmetry to the Poincare one,
with the emergent NG boson  associated with  graviton.
The latter paradigm  proves to be nicely fitted  to gravity, 
naturally incorporating the generic signatures  of GR:
the symmetric second-rank tensor field possessing the self-interactions
through  a derivative in the  ratio to   the Planck mass treated  as
a scale of the affine SSB.
By means of  the group-theoretic techniques of  the nonlinear realizations 
(NRs) and   nonlinear models (NMs) for  SSB~\cite{Cole}--\cite{Ish} 
this  allows   to justify  GR  as the metric theory of gravity, 
with the graviton being ultimately  a Goldstone  field.
The question then arises: to what extent such an alternative approach to gravity 
may be of use for  going beyond GR, as was advocated above.\footnote{For 
 gravity as  an affine-Goldstone phenomenon 
and beyond, cf.,~\cite{Pir3}.}$^,$\footnote{For  graviton  
as a Goldstone field in the  Lorentz gauge theory, cf.,~\cite{Sard}.}   
 
To this end, in the present paper we 
merge  the two above-mentioned routes of the GR modification,
so to say,  ``in width'' and ``in depth''.
While  the quartet-modified GR (like  
GR itself and its direct siblings) is originally 
based upon the  pseudo-Riemannian/metric paradigm in the 
conventional EFT  frameworks, 
 we are  now going  to consider a more advanced  EFT
of gravity (encompassing the plain  one) 
based upon the group-theoretic/NG paradigm 
in  the   NM frameworks. 
At that, due to  the internal and external symmetries in the case at hand
coinciding we construct such an NM
for consistency  in the two steps.
In Sec.~2,  in an auxiliary  affine space, 
a progenitor of the Lorentz  spacetime, we start by  presenting 
NRs for the spontaneously broken global affine symmetry and  
then build  an affine-Goldstone  NM.  
In Sec.~3, such a  construction is proliferated    
to an arbitrary  world  manifold, a progenitor  
of the pseudo-Riemannian spacetime. 
It is shown  that NM so constructed   supersedes  the earlier 
introduced EFT of the quartet-metric  gravity,  with a  clarification 
and elaboration of the latter. For illustration, the two 
extreme  cases of the quartet-metric  gravity  are  exposed in more detail.
In Summary,  there are collected  the main advantages of presenting  
the quartet-metric gravity 
as NM in the group-theoretic/NG paradigm  
vs.\  EFT in the pseudo-Riemannian/metric paradigm.  
Finally,  we  point out towards a conceivable prospects of going  beyond 
the NM frameworks  above  the Planck scale  towards an underlying theory  
for  the emergent  gravity and  spacetime.
All the  constructions  for the affine-Goldstone NM  are  presented  throughout 
in a nutshell directly adopted to the case  at hand.
For completeness,  in Appendix  
there  are exposed in a more generality 
the group-theoretic basics concerning  SSB, NRs and NMs.

\section{Affine-Goldstone NM}

\subsection{ Affine-to-Poincare SSB} 

Let us first indicate  the basic results 
in the affine space for NM  to be used  subsequently in  constructing the 
modified gravity in the world manifold.\footnote{Taking these results 
for granted one may directly consider their application in Sec.~3.}
To start, let  $R^4= \{z^a \}$, be  an auxiliary   
affine space,\footnote{Such an affine realization space   may  physically be 
thought of as corresponding to  a putative 
primary metricless world  (or  a ``bare''  vacuum).} 
with  the  coordinates $z^a$ marked by 
the  indices $ a,b, \dots= 0,\dots,3$,  acted upon 
by  the  global affine transformations 
$(A,C)\in IGL(4,R)\equiv Aff(R^4)$ as  $z\to A z +C$, 
or in the full notation
\be
(A,C) \ : \ z^a\to A^a{}_b z^b + C^a.
\ee
At that,  the index $0$ is {\em a priori} nothing but  
a notation.
The NM  under construction is  based upon  
the  SSB  pattern\footnote{In fact, we  consider the 
full  affine and inhomogeneous Lorentz (Poincare) groups. 
But remaining  unbroken,  the inhomogeneous parts  of the  groups  
are explicitly omitted.}
\be \label{patt}
G\equiv  GL(4,R)\to H\equiv  SO(1,3),
 \ee
 with 
$SO(1,3) \subset GL(4,R)  $  the  residual  (global) Lorentz
subgroup. Remaining  unbroken, the latter serves as a classification group 
for the field representations.
Just after  a particular embedding of  the unbroken
subgroup $SO(1,3)$ into the broken  $GL(4,R)$ the affine space 
gets converted  into the Lorentz  spacetime, 
with a foliation onto the time and space:  $z^{a}=(z^0, z^i)$,
$i=1,2,3$, where  the affine index $0$ is assumed to be chosen 
so as to  acquire its  conventional time  meaning.
The respective  NM  $G/H=  GL(4,R)/SO(1,3) $
describes the  SSB $G\to H$, with 
the appearance of the $d _{G/H}   =d_G- d_H    =  10 $ 
NG bosons  associated ultimately with the (tensor) graviton 
and  the additional gravitational dark components.\footnote{Note
that all the consideration 
is technically   insensitive to the dimension 
of spacetime $d\ge 3$, as well as  to  
its signature $(p,q)$, $p+q=d$. 
For this reason, the quartet-metric gravity, containing in metric the 
gravity components   in excess of  the  tensor graviton, 
may rather be called the multi-component gravity.}

\subsection{Quasi-symmetric  NR} 

 The construction of the proper NM, call it  
the  affine-Goldstone one,  follows 
(with slight modifications)  to the general procedure  
 of NRs (see, e.g., Appendix).
 What is of specifics   in the case at hand 
 is that here the (left) coset representative $\mathring \vart$,  associated 
 with the  NG field, 
may be parametrized as a $4\times 4$ (local) matrix $\mathring\vart^a_{(b)}(z)$  
transforming under  $A\in GL(4,R)$ as  
$\mathring\vart\to A\mathring\vart \La^{-1}(A,\mathring\vart)$. 
or in the full notation 
\be 
A \ : \   \mathring\vart^a_{(b)}(z) \to  \mathring \vart'^a_{(b)}(A  z)  =  A^a{}_c 
\mathring\vart ^c_{(d)} (z)\La^{-1d }{}_{b}(A,\mathring\vart(z)),  
\ee 
with both the field $\mathring \vart$ and 
the coordinate $z$ transforming under $A$.
At that, $\La=\La(A,\mathring\vart)\in SO(1,3)$ 
is to be properly defined  
(henceforth, the term nonlinear realization)
to specify a particular  NR.  
In what is shown  above and  what follows, an affine index 
in the parentheses indicates that it undergoes 
 transformations just  under the
residual Lorentz subgroup $SO(1,3)$. Only such   
indices  are allowed to be raised and lowered
 by means of the Lorentz-invariant Minkowski symbol $\eta^{ab}$ (and, 
respectively, $\eta_{a b}$) without an explicit 
violation of  the affine symmetry.
With $\mathring\vart^a_{(b)}$ representing a group element 
it has an inverse $\mathring\vart^{(a)}_b$ 
transforming under $A\in GL(4,R)$, in short, 
as\footnote{We 
designate  the  inverse  matrix $\mathring\vart^{(a)}_b$ 
in a short-hand notation 
as  $\mathring\vart^{-1}$, etc.}  
$\mathring\vart^{-1}\to\La(A,\mathring\vart) \mathring\vart^{-1} A^{-1}$.

An arbitrary   $4\times 4$  matrix  contains
{\em a priori}  sixteen components.   
 To  fix a (left) coset representative  $\mathring\vart$ there should be 
 leaved   in the matrix  just ten independent  components,
equal to the difference of the dimensions of the linear and Lorentz groups, 
through  imposing  six auxiliary  conditions.  
In the case at hand 
the most natural  choice is the   quasi-symmetric  condition, in short,    
$\mathring\vart\eta =(\mathring\vart\eta)^T= \eta\mathring\vart^T$, or
\be 
\mathring\vart^{a}_{(c)}\eta^{cb} =  \mathring\vart^{b}_{(c)} \eta^{ca},
\ee  
eliminating the (quasi)-anti-symmetric part 
present in  the arbitrary $4\times 4$ matrix. 
With account for  
the  pseudo-orthogonality property of $\La\in SO(1,3)$, 
in short, $ \eta \La^T\eta=\La^{-1}$, or 
\be
 \eta^{a c} \La^{ d}{}_{ c}\eta_{db}=    \La^{-1 a}{}_{ b} ,
\ee
to preserve  the  imposed  condition  under  an arbitrary $A$ 
there  should fulfill
\be\label{NRcond}
A\mathring\vart\eta \La^{T}=  \La \mathring \vart \eta  A^T .
\ee
Under restriction by  $A=\La_0$, with  
an arbitrary global  $\La_0\in SO(1,3)$, 
we clearly get $\La=\La_0$ implying, in particular, $\La=I$ at $ A=I$, 
with $\mathring \vart$ remaining unchanged.
A general  solution to eq.~(\ref{NRcond})  may be looked for  by perturbations, 
uniquely  at least in a vicinity of unity.
The resulting   $\La=\La(A,\mathring\vart)$ defines the particular NR,
which reduces to   the linear representation 
when restricted  by  the unbroken Lorentz subgroup;
\be
\La_0 \ : \ \mathring\vart \to \La_0\mathring\vart \La_0^{-1},
\ee
or,  with account for the pseudo-orthogonality 
of $\La_0$, as the symmetric Lorentz tensor:
\be
\La_0 \ :  \mathring\vart\eta \to \La_0\mathring\vart\eta  \La_0^{T}. 
 \ee
At last, decomposing  $\mathring\vart$  in the weak-field limit as 
\be\label{GB}
\mathring\vart^a_ {(c)} \eta^{cb}\simeq \eta^{ a b}+ 
\chi^{a  b},
\ee
with the symmetry  $\chi^{a  b}=\chi^{b a}$ being preserved 
by  the classification Lorentz subgroup, 
we  may interpret~$\chi$ as a Lorentz-tensor NG boson for SSB at hand.  

Imposing other auxiliary conditions  we may  similarly get other particular NRs,
all of them  being  equivalent   (within the full NM).
Dealing exclusively with a  particular NR  may be  cumbersome.  
For this reason, we get rid of the explicit auxiliary conditions  and 
consider below  (in accord with the general theory of NRs) 
a linearized realization  explicitly equivalent to any particular NR,
but being, in fact, much simpler.

\subsection{Local-Lorentz  linearization}

Instead of a (left) coset representative $\mathring\vart(z)$, let us now   choose 
as a field variable  for SSB $G\to H$
the (left) coset itself, i.e., the whole subset (not, generally, a subgroup) 
of the elements of $G$ equivalent {\em modulo}
the (right) multiplication by $H$ 
and  containing $\mathring\vart$ as a representative: 
$\vart \equiv  \{\mathring  \vart \La^{-1}\}$, $\La\in H=SO(1,3)$.
To this end,  define for the case at hand   as a new field variable  
the $4\times 4$ local matrix $\vart^a_\al(z)$ (with an inverse  $\vart^\al_ a$)
transforming under $A\in GL(4,R)$ up to 
an arbitrary  $\La(z) \in SO(1,3)_{\rm loc}$, in short, as  $\vart \to A\vart \La^{-1}(z)$, or 
\be\label{ALaz}
(A,\La(z)) \ :  \     \vart^a_{\al}(z) \to   
\vart'^a_{\al}(Az) =  A^a{}_b \vart ^b_{\beta}(z) \La^{-1\beta }{}_{\al}(z),
\ee
(respectively, $\vart^{-1}\to \La(z)\vart ^{-1} A^{-1}$).
In what is shown  above,  $SO(1,3)_{\rm loc}$ is 
an auxiliary local Lorentz group 
(not a subgroup of $GL(4,R)$ as before).
At that, though   $\vart^a_{\al}(z)$ contains 
formally sixteen components,  six of them 
can be eliminated  by means of the 
auxiliary transformations $\La(z)\in SO(1,3)_{\rm loc}$ leaving  
as required  precisely ten independent components.
The remaining   components may ultimately 
be  associated  with   the NG boson  
arising under  SSB $GL(4,R)\to SO(1,3)$.
Namely, fixing properly a  gauge for $SO(1,3)_{\rm loc}$ 
we can get any particular NR.
Thus, transforming 
$\vart\to \vart \mathring \La^{-1}$, with $\mathring\La\in SO(1,3)_{\rm loc}$ 
satisfying, with account for the pseudo-symmetricity 
of $\mathring \La$,  to the relation, in short,  
$\vart\eta \mathring\La^{T} 
=  (\vart\eta \mathring\La^{T})^T  = \mathring\La \eta\vart^T$, or
\be
\vart^a_\al \eta^{\al\ga} \mathring\La{}^{\beta}_\ga|_{\beta=b}=
\vart^b_\beta \eta^{\beta\ga} \mathring\La^{ \al}{}_\ga|_{\al=a},
\ee
we can achieve  that 
$\mathring \vart  \equiv \vart\mathring  \La^{-1}$ 
gets quasi-symmetric, $\mathring\vart\eta= (\mathring\vart\eta )^T=\eta \mathring\vart^T $,
recovering thus the quasi-symmetric NR.
In the same vein, we can get any other particular NR, 
all of them (in accord with the general NR theory) 
being equivalent  to each other
and to the local-Lorentz linearized realization.

In what follows, we will use such a realization, with  
the  symmetry group  in the factorized affine-Lorentz 
form:\footnote{At that, such an  auxiliary 
non-compact local-Lorentz   group $SO(1,3)_{\rm loc}$, in distinction 
with  a putative gauge one,  is not  equipped with  the  physical gauge fields, 
so that in the case at hand  there appears  no apparent problems with 
unitarity.}
\be\label{GHloc}
G\times H_{\rm loc}= GL(4,R)\times SO(1,3)_{\rm loc}.
\ee
The  SSB  pattern eq.~(\ref{patt}) and the hidden 
local symmetry in eq.~(\ref{GHloc}) 
may be treated  as an essence  of gravity,
succinctly  encompassing the bulk of the gravity appearance,
{\em modulo} the  world environment 
(see, Sec~3).
According to eq.~(\ref{GHloc}), there  
can be envisaged the three generic  types 
of the (finite dimensional) affine-Lorentz fields: 
the affine protometric, the affine tensors 
and Lorentz spinors, to be  specified  below.

\subsection{Affine-Lorentz  fields}

\subsubsection{Affine protometric}

To describe the dynamics of the affine NG field $\vart$ 
consider, in accord with the general prescription, 
the (slightly modified)  so-called Maurer-Cartan form, in short, 
$\Omega_a= \vart^{-1} \partial_a\vart\eta$, or  
\be
\Omega_c^{\al\beta}= 
\vart_d^{\al} \partial_c\vart^d_{\ga} \eta^{\ga\beta}. 
\ee
According to eq.~(\ref{ALaz}),   $\Omega_a$ is a vector  under $GL(4,R)$  
and, with account for the pseudo-orthogonality of $\La$, transforms 
under $SO(1,3)_{\rm loc}$ inhomogeneously~as 
\be\label{inhom}
(A,\La(z)) \ : \ \Omega_c(z)  \to  \Omega'_c(Az)  
= A^{-1b}{}_c\Big(\La \Omega _b\La^T +\La\partial_b \eta \La^T\Big).
\ee
Decomposing  $\Omega_c $  in the symmetric and anti-symmetric parts, in short, 
$\Omega^\pm_{c}\equiv (\Omega_c\pm \Omega_c^T)/2$, or 
\be
 \Omega^{\pm\al\beta}_{c}\equiv 
 \frac{1}{2}(\Omega_c^{\al\beta}\pm \Omega_c^{\beta\al}),
 \ee
we can see from eq.~(\ref{inhom}) that  the symmetric part 
transforms homogeneously under $\La(z)$,
while  the anti-symmetric part transforms inhomogeneously.
This allows to  use these parts   separately.
Namely, the inhomogeneously transforming anti-symmetric  part  $\Omega_c^{-}$
may serve as a  Lorentz connection in the case of the fermion  matter fields
(see, later on). 
On the other hand, the symmetric  part  
$\Omega_c^{+}$  being a  Lorentz tensor 
may   be used to describe the NG field 
$\vart$ by its own   through constructing  the arbitrary 
local-Lorentz invariant combinations.

Otherwise, due to a freedom 
(within the full  NM) of choosing the field variables,   
 to facilitate the procedure we may  
 start directly from the local-Lorentz invariant combination in the short-hand notation  
 $ \theta=\vart\eta \vart^T$, $\theta=\theta^T$, or
 \be
\theta^{ab}\equiv \vart^a_{\al}  \eta^{\al\beta}  \vart^b_{\beta},  
\ \ \theta^{ab}=\theta^{ba}, 
\ee
with the inverse
$ \theta^{-1}=\vart^{-1T}\eta \vart^{-1}$, 
 $\theta^{-1}=\theta^{-1T} $,  or\footnote{To distinguish 
 from  the affine-violating combination
$  \eta_{ac}\eta_{bd}\theta^{cd}\neq \theta_{ab}$, 
or  in short $\eta\theta\eta\neq \theta^{-1}$.} 
\be
\theta_{ab}\equiv
\vart^ \al_a  \eta_{\al\beta} \vart^\beta_b, \ \ \theta_{ab}=\theta_{ba}. 
\ee
 So, instead of the affine-Lorentz bi-vector  $\vart^a_\al$, corresponding 
to the affine NG boson, let us  consider as a  new   field variable 
the  genuinely-affine tensor $\theta^{ab}$ (or its inverse $\theta_{ab}$)
corresponding to the point-like   correlated symmetric pair of the NG bosons.
Being  symmetric, $\theta=\theta^T$, this tensor  automatically contains
the required  number, ten,  of the independent components 
irrespective  of the particular gauge for  $\La(z)$. 
Under $GL(4,R)\times  SO(1,3)_{\rm loc}$
the field $\theta^{ab}$ transforms,  in short,~as  
\be
(A,\La(z)) \ : \ \theta(z)\to  \theta'(Az) =   A\theta(z) A^T,
\ee
with the symmetry of $\theta$  automatically preserved. 
As a result,  NR for such a  field   becomes 
the conventional linear representation exclusively  of the 
affine group. 
In the weak-field limit we have 
\be
\theta^{ab}\simeq \eta^{ab}+2 \chi^{ab}, 
\ee
with the NG boson  $\chi^{ab}$ from eq.~(\ref{GB}).
Further,  instead of a derivative of $ \theta^{ab}$ we can equivalently use
a genuinely-affine  tensor $\Ga^c_{ab}$ 
defined as\footnote{Clearly,  
the  symmetric affine  tensor 
$\Ga^c_{ab} = \Ga^c_{ba}  $, like $\partial_c \theta_{ab}$,  
contains  the same number, forty,  of components as 
the symmetric affine-Lorentz tensor  $\Omega_a^{+\al\beta}$, 
allowing ultimately to substitute each other.} 
\be\label{Ga}
\Ga^c_{ab}  \equiv \frac{1}{2}\theta^{cd}
(\partial_a \theta_{bd}+ \partial_b
\theta_{ad} - \partial_d \theta_{ab}),
\ee
so that inversely 
\be
\partial_c \theta_{ab}=  \theta_{ad}  \Ga_{bc}^d +
\theta_{bd}\Ga_{ac}^d .
\ee
Looking at the first sight optional, the affine tensor $\Ga^c_{ab}$ 
proves to be of a principle importance for the geometrical interpretation 
of the  affine-Goldstone NM.
The correlated pair of the NG fields,  
$\vart\eta\vart^T$ (resulting ultimately 
in the pseudo-Riemannian metric)  may be called the affine 
protometric.
Such a field being   self-sufficient to construct
 in the affine space  an   NM by its own,
 proves at the same time to be inevitable to  consistently 
incorporate   other  fields considered below.

\subsubsection{Affine tensor fields}  

Consider now an arbitrary affine-tensor field 
$\Phi^{a_1,. . . }_{b_1,\dots}(z)$ transforming under $A\in GL(4,R)$ 
independently in each of the upper and lower 
indices through $A$ and $A^{-1T}$,  respectively. 
The field   $\theta^{ab}$ (or $\theta_{ab}$) 
may serve  as a counterpart of the metric  appearing  due to  
the affine SSB in the originally  metricless affine space.
It allows to raise and lower the 
affine tensor indices to construct the   bi-linear  affine scalars, etc.
E.g., by means of $\theta_{ab}$ we can construct from, say,  
an affine vector $U^a$ a quadratic combination  $U^a U^b \theta_{ab}$
serving as an  affine scalar, etc.

To facilitate the manipulations with such and similar constructions,
in particular with  their partial derivatives $\partial_a$,  we can 
define in terms of the previously introduced affine  tensor  $\Ga^c_{ab}(\theta)$ 
a counterpart of the  covariant derivative. 
To this end, let us define the basic affine  covariant derivative $\na_a$
for some vector fields $V_a$ and $U^a$  as follows:
\bea
\na_a U^c&=&    \partial_a U^c+ \Ga^c_{ad}U^d,\nn\\
\na_a V_b&=&  \partial_a V_b - \Ga^d_{ab}V_d ,  
\eea
so that 
$\na_a (U^c V_c)=\partial_a (U^c V_c)$, etc.  
Likewise, we can further proliferate  the action of $\na_a$ on an  arbitrary 
affine tensor $\Phi^{c_1, ...}_{b_1,...}$ 
through the independent action  on each of the indices  as shown above.
It follows hereof that 
\be\label{conserv}
\na_a \theta^{bc}= \na_a \theta_{bc}=0,
\ee
so that    $ \na_a V^b \equiv   \na_a \theta^{bc} V_c =\theta^{bc}\na_a V_c$, 
etc.
This allows, in particular,  to disentangle the operations of differentiation  and those  
of raising/lowering the affine  indices.
Likewise, we can deal with the  arbitrary affine-tensor fields and 
their combinations.\footnote{The property (\ref{conserv})  is, 
by construction,  peculiar  to $\theta_{ab}$.  
So that an arbitrary  symmetric affine tensor 
$T^{ab}$ formally similar to  $\theta^{ab}$, generally,  fulfill   $\na_a T^{ab}\neq 0$.}
The affine tensor $\Ga^c_{ab}$ may thus be termed as 
the  affine protoconnection.\footnote{Moreover,  $\na_a$  allows to 
consider  in the affine space the arbitrary curvilinear coordinates, 
with $\Ga^c_{ab}$ serving for such a purpose 
as a genuine affine connection.}

To account  for the 
tensor matter fields and the continuous media 
it is thus sufficient  to  consider just the correlated pair $\theta=\vart \eta \vart^T$ 
of the affine NG fields,  
with transformations  only under $A\in GL(4,R)$.
But  to include  the fermion matter fields  
it is necessary to take into account 
the  affine NG field $\vart$ itself  as the basic variable,  
with transformations also under  $\La(z)\in SO(1,3)_{\rm loc}$ 
to be considered below.

\subsubsection{Lorentz spinor fields} 

Let $\rho(\La)$ be a finite-dimensional linear representation of the 
Lorentz group,\footnote{Note that 
in distinction with the affine tensors,
the finite-dimensional spinors make sense   only 
at the level of the residual Lorentz symmetry after the SSB $GL(4,R)\to SO(1,3)$. 
Thus the conventional matter fields may appear only after the affine SSB.}   
with a generic fermion field
 $\Psi$ transforming 
under $\La^{\al\beta}= \eta^{\al\beta} 
+ \la^{\al\beta} +{\cal O}(\la^2)$ as
\be
\La (z)\   : \  \Psi\to       \Psi'=   \rho(\La)\Psi= 
\Big( I+\la^{\al\beta}(z) \rho(L_{\al\beta})   \Big ) \Psi  +{\cal O}(\la^2) , 
\ee
where   $\la^{\al\beta} = -\la^{\beta\al}$, $|\la|\ll 1$, are the infinitesimal (local) parameters
and  $L_{\al\beta}=- L_{\beta\al }$  are   the generators of the Lorentz 
group.
Now, the Lorentz anti-symmetric part $\Omega^{-\al\beta}_a$, to be called the  spin-connection,
allows to introduce a covariant under  $SO(1,3)_{\rm loc}$ derivative~as 
\be
\na_a  \Psi=\partial_a \Psi +\Omega^{-\al\beta}_a \rho(L_{\al\beta})\Psi,
\ee
where   $\Omega^{-\al\beta}_a$ 
plays a role of the auxiliary  (composed) Lorentz gauge field
transforming according to eq.~(\ref{inhom})  as
\be
\La(z) \ : \ \Omega^{-\al\beta}_c \to  \Omega'^{-\al\beta}_c =     \Omega^{-\al\beta}_c       
+\la^\al{}_\ga \Omega^{-\ga\beta}_c - \la^\beta{}_\ga \Omega^{-\ga\al}_c
-\partial_c \la^{\al\beta}+{\cal O}(\la^2).  
\ee 
It follows from what is shown above and  
the commutation relations for $L_{\al\beta}$ that $\na_a\Psi $  
transforms  under the infinitesimal $\La(z)$ homogeneously like $\Psi$ itself: 
\be
\La(z) \   : \  \na_a \Psi \to \na'_a \Psi'\equiv   
\na'_a \Big(\rho(\La) \Psi\Big) =\rho(\La)  \Big(\na_a  \Psi\Big).
\ee
The same properties fulfill for a finite $\La(z)$.             

To illustrate, for a Dirac bi-spinor $\psi$  we can choose   
 the Lorentz generators  as $\rho(L_{\al\beta})\sim\si_{\al\beta}\equiv [\ga_\al,\ga_\beta]/2$,
with the  Lorentz $\ga$-matrices conventionally defined by  the anti-commu\-tation relations 
\be
\{\ga^{ \al},\ga^{ \beta}\} =2  \eta^{\al\beta}.  
\ee
At that, the affine $\ga$-matrices, $\ga^a\equiv \vart_{\al}^a\ga^{\al}$ 
(respectively, $\ga_a=\theta_{ab}\ga^b=\vart^{\al}_a\eta_{\al\beta}\ga^{\beta}$),
to be used in constructing the affine-invariant combinations,
fulfill the relations
\be
\{\ga^{a},\ga^{ b}\} =2   \theta^{a b}.
\ee
Likewise, we can define the affine matrix 
$\si_{ab}\equiv  \vart^{\al}_a \vart^{\beta}_b \si_{\al\beta}= [\ga_a,\ga_b]/2$, etc.
Such the affine matrices are to  be used in constructing  the Lorentz-invariant 
bi-linear  combinations transforming only under the affine group, say,   $\bar\psi \ga_a\psi$  
as the affine vector,    $\bar\psi \ga^a \na_a\psi$ as the affine scalar,     $\bar\psi \si_{ab}\psi$   
as the affine tensor,  etc. More generally, we can consider the mixed 
affine-Lorentz spin-tensors $\chi^{a_1, . . .}_ {b_1, . . .} $ to construct from them 
the invariant combinations both under $GL(4,R)$ 
and $SO(1,3)_{\rm loc}$.

\subsection{Second-derivative affine protogravity}

Now we are in a position to construct 
the affine-Goldstone NM $G/H=GL(4,R)/SO(1,3)$.
First, we do this  in the affine realization 
space and then  embed  the construction  into 
the physical world manifold converting thus  
the latter into  the pseudo-Riemannian spacetime.
We restrict the subsequent consideration exclusively by the  NG 
part of the action in terms of the protometric. 
The  inclusion of the tensor and spinor matter  fields
is, in principle,   straightforward 
by means of  the techniques presented above.
In the case of just the spontaneous 
breaking  of the affine symmetry (without its explicit violation),   
the  affine NG  
part of NM can  depend only on $\theta^{ab}$ and its derivatives.
Allowing for an explicit violation of the affine symmetry to the Lorentz one, 
NM may  also include the Minkowski symbol $\eta_{ab}$ (or $\eta^{ab}$).
By this token, the NG boson    becomes, in fact, 
the  pseudo-Nambu-Goldstone (pNG) one.

Thus, the most general Lorentz   scalar Lagrangian  
for the affine pNG boson  may generically be presented  as 
\be
L_G=L_G(\partial_c\theta^{ab}, \theta^{ab},\eta_{ab}).
\ee
To account for $\eta_{ab}$,   we may  equivalently consider the Lorentz tensor 
\be
{H}^a{}_b\equiv \theta^{ac}\eta_{cb}, 
\ee
(so that $ \eta_{ab}=\theta_{ac} {H}^c{}_b$)
as well as its inverse $H^{-1}{}^b{}_a\equiv \eta^{bc}\theta_{ca}$,  
explicitly  violating the affine symmetry to the Lorentz one.
With the derivatives $\partial_c \theta_{ab}$ expressed 
through $\Ga^c_{ab}$, such a  Lagrangian may generically  be partitioned  
into  the kinetic and potential contributions as follows:
\be
L_G=\ka_0^2 K( \Ga^c_{ab},  \theta^{ab},{H}^a{}_b)  - V({H}^a{}_b), 
\ee
where the parameter  $\kappa_0$  of the dimension of mass designates  an SSB scale.
The  kinetic term 
in the second-derivative  order with the minimal explicit violation
of the affine symmetry is  as follows:
\be\label{K} 
K =  \frac{1}{2}\sum_{p=1}^5   \ve_p({H}) {K}_p (\Ga^c _{ab},  \theta_{ab} ),
\ee
where $\ve_p$,  $p=1,\dots, 5$, are some  free dimensionless  parameters,  
generally, dependent on $H$, with  
the partial kinetic contributions 
\bea\label{O_p}
 {K}_1= \theta^{ab} \Ga_{a c}^c  \Ga_{bd}^d,   && 
 {K}_2=\theta_{ab}  \theta^{cd}\theta^{ef}
 \Ga_{cd}^a    \Ga_{ef}^b,\nn\\
 {K}_{3}=\theta^{ab}  \Ga_{ab}^c \Ga_{cd}^d.      &&   
{K}_4 = \theta_{ab}  \theta^{cd}  \theta^{ef}  \Ga_{ce}^a
 \Ga_{df}^b    ,\nn\\  
{K}_5=  \theta^{ab}   \Ga_{ac}^d   \Ga_{bd}^c , 
 \eea
being independent of $\eta_{ab}$.
Further, by admitting  the  second-order derivatives  of $\theta$ 
and extending formally  the affine symmetry  to that 
under  the arbitrary curvilinear coordinate 
transformations  in the affine space,\footnote{To 
generate   such  transformations  in the affine space it is, in fact, sufficient
to consider  a  closure 
of the  affine  and  conformal transformations~\cite{Ogi2}.}
we can formally construct   from the protometric $\theta_{ab}$ 
a counterpart of the Riemannian  tensor,
then a Ricci tensor $R_{ab}$ and, at last, 
a Ricci scalar $R\equiv \theta^{ab} R_{ab} $.
By this token, we can supplement  $L_G$ by  the term\footnote{In principle, 
$R$  may be expressed, up to a surface contribution,
through a  linear combination of $K_3$ and $K_5$ 
(or rather, say, $K_3$ may be expressed through $K_5$ and $R$, etc,).
Not having a prior preference, we leave explicitly all the terms shown above.} 
\be
L_g\equiv- \frac{1}{2}\ka_0^2 (1+ \ve_0({H})) R,
\ee  
with a  dimensionless parameter $\ve_0$.
And finally, the potential term $V(H)$ is an 
arbitrary Lorentz-scalar  polynomial of ${H}^a{}_b$,
like $\mbox{\rm tr} (H^2)\equiv {H}^a{}_b{{H}}^b{}_a$, etc., supplemented,  
generally, by $\det(H^a{}_b)=\det(\theta^{ab}) \det(\eta_{ab})  $, as well as  a  
constant part $V_0$.

Likewise, we could  construct  
in the   affine realization space, a progenitor of the Lorentz spacetime,  
the most general affine and locally-Lorentz invariant NM
for the pNG boson supplemented by the matter fields.
Nevertheless, such a  construction 
is to  be considered as no more than  an  auxiliary one.
To describe the real  world,  the construction  should  be 
 proliferated   onto an arbitrary  world  manifold, 
 a progenitor of the pseudo-Riemannian  spacetime, 
to be considered below.

\section{Quartet-metric  EFT and beyond}

\subsection{General covariance}

Let  now the real world -- a set of the primary events (points) -- 
be modeled by a four-dimensional topological manifold $M^4$ endowed with 
the arbitrary smooth enough  coordinates $x^\mu$, $\mu=0, \dots, 3$.
(As before, the index $0$ is originally nothing but a notation.)
Let  $M^4\leftrightarrow R^4$ be a local one-to-one mapping of the   manifold onto  
the  affine realization space by means of   some invertible 
transformation  functions $z^a=Z^a (x)$,
with $x^\mu=x^\mu(z)$. Call  the so-constructed  distinct coordinates  $z^a =Z^a(x)$
by definition  the quasi-affine ones on $M^4$. Generally, 
 the  reversibility of the mapping may  be 
hampered by  some singular subsets of  $M^4$,
with the mapping being in fact patch-wise.
In terms of $z^a$, let us map  all the quantities  on $R^4$ into  the 
respective quantities  on $M^4$.
{\em A~priori}, the mapping given by $Z^a(x)$ could remain  unspecified
signifying ultimately an incompleteness of the approach.  
To eliminate this uncertainty
we treat $Z^a(x)$  by construction  as the dynamical variables 
describing gravity on par with $\theta_{ab}$.
Otherwise, this means that the   gravity is a net result of imposing 
both  the local field-theoretic (due to $\theta_{ab}$) and geometrical   
(due to $Z^a$)  effects.
For the general covariance (GC) let us express  further the so-obtained structures   
in terms of the  arbitrary observer's coordinates $x^\mu =x^\mu(z)$.
Introducing  for such the transformations on $M^4$  the  
respective quasi-affine  tetrad 
${\zeta}{}_\mu^a(x)\equiv \partial_\mu Z^a$ and its  inverse,
${\zeta}{}^\mu_a (x)\equiv\partial x^\mu/\partial  z^a|_{z=Z(x)}$,  we can proliferate 
NM  from the affine realization space in the GC manner onto the world manifold 
converting the latter into the pseudo-Riemannian one.
Operationally, this can be achieved from the preceding results through 
the coordinate substitution $z^a \to x^\mu$ followed
by the proper substitutions of the basic quantities,  
which may be partitioned  in the three groups as shown below.

\subsubsection{Enhanced  contributions} 

The first  group of substitutions  is  as follows:
\bea\label{subst}
\theta_{ab}&\to&g_{\mu\nu}\equiv {\zeta}{}_\mu^a {\zeta}{}_\nu^b \theta_{ab} ,
\nn\\\
\theta^{ab}&\to&g^{\mu\nu}\equiv {\zeta}{}^\mu_a {\zeta}{}^\nu_b \theta^{ab} , \nn\\
R_{ab}&\to&R_{\mu\nu}\equiv {\zeta}{}_\mu^a {\zeta}{}_\nu^b  R_{ab},\nn\\
R\equiv \theta^{ab} R_{ab}&\to&R\equiv g^{\mu\nu} R_{\mu\nu} ,\nn\\
\det(\theta_{ab})&\to&g\equiv \det(g_{\mu\nu})=
\det({\zeta}{}^a_\mu)^2\det(\theta_{ab}),\nn\\  
\ga^a&\to& \ga^\mu\equiv {\zeta}{}_a^\mu\ga^a,
\eea
with  
\be
\{\ga^\mu,\ga^\nu\}= 2g^{\mu\nu }.
\ee
In what is shown  above, the NG field in disguise,  
$g_{\mu\nu}$, is to be treated as 
the pseudo-Riemannian  metric 
with the inverse $  g^{\mu\nu}$, 
while $R_{\mu\nu}$ proves to be  the conventional Ricci curvature tensor 
constructed from $g_{\mu\nu}$.
This group of term is relevant for GR and its direct siblings.

Otherwise, the quantities above   may  be expressed  through a 
local-Lorentz tetrad $\vart^\al_\mu$ defined~as  
\be
\vart^\al_a\to   \vart^\al_\mu  \equiv {\zeta}{}^a_\mu\vart^\al_a,
\ee
so that, say,
\bea
g_{\mu\nu}&=&\vart^\al_\mu \vart^\beta_\nu\eta_{\al\beta}, \nn\\
\sqrt{-g}&=& \det (\vart^\al_\mu) (-\det(\eta_{ab}))^{1/2}   ,\nn\\
\ga^\mu&=&\vart_\al^\mu \ga^\al, 
\eea
etc. In principle, the local-Lorentz tetrad $\vart^\al_\mu$ may be chosen 
as an alternative to the metric $g_{\mu\nu}$.
At that, the   tetrad 
$\vart^\al_\mu(X)$ in an arbitrary  chosen point $X$
may be  associated  with  the so-called 
locally-inertial tetrad $\hat \vart^{\hat\al}_\mu(X)$ in GR   
 through a  locally-inertial  gauge  given   
 by$\hat\La(X)$  as
 \be
\hat \La^\al{}_\beta (X)\vart^\beta_\mu(X)|_{\al=\hat\al}=
\hat \vart^{\hat \al}_\mu(X)  
\equiv \partial \hat x_X^{\hat \al}/\partial x^\mu|_{x=X},
\ee
 were $ \hat x_X^{\hat \al}$ are the locally-inertial 
in  a vicinity of the  point $X$ 
coordinates.\footnote{Remind that such   coordinates      
are defined as those wherein 
the metric is locally-flat:  
$\hat g_{\hat\al\hat\beta}(\hat x_X)|_{\hat x_X\simeq X}
\simeq \eta_{\hat\al\hat \beta}$ 
up to the quadratic deviations of $\hat x_X$ from $X$.}  
Due to the local-Lorentz symmetry  using the 
locally-inertial coordinates to describe fermions 
is not as crucial  as in the plain GR.
For GR and its direct siblings, the dependence on $Z^a$
gets  completely hidden, with such the theories becoming 
the exclusively  metric ones.\footnote{Hence, restricting  
{\em ab initio} NM by 
the pseudo-Riemannian manifold $M^4$~\cite{Ish} 
we  could in the end  recover only the part 
of NM corresponding to GR (and its direct siblings).
For this reason,  the two-stage  realization of NM to get  
the quartet-modified GR  is paramount.}$^,$\footnote{The respective  
group of terms originates  ultimately from those 
in the affine realization space
with the  affine symmetry
enhanced to the conformal one, as it was mentioned previously.}

The quasi-affine coordinates  manifests themselves  only for the GR extensions 
originating from terms in the affine realization space with the normal 
or suppressed strength. The proper terms   correspond to 
the non-enhanced affine symmetry,
respectively, without or with an   explicit violation  of the affine symmetry 
to the Lorentz one, presented  below.

\subsubsection{Normal contributions}  

The second group of terms consists of  the connection-like  GC tensor
\be\label{B}
\Ga^c_{ab}\to B^\la_{\mu\nu}\equiv  \Ga^\la_{\mu\nu}-  \ga^\la_{\mu\nu},
\ee
where
\be\label{Ga}
\Ga^\la_{\mu\nu}\equiv {\zeta}{}^\la_c {\zeta}{}_\mu^a {\zeta}{}_\nu^b \Ga^c_{ab}
=\frac{1}{2}g^{\la\ka}(\partial_\mu g_{\nu\ka}+ \partial_\nu
g_{\mu\ka} -
\partial_\ka g_{\mu\nu})
\ee 
is nothing but the Christoffel connection 
for  the pseudo-Riemannian metric  $g_{\mu\nu}$ and 
\be\label{ga}
 \ga^\la_{\mu\nu}\equiv {\zeta}{}^\la_a\partial_\mu {\zeta}{}^a_\nu = {\zeta}{}^\la_a\partial_\nu {\zeta}{}^a_\mu
 \ee
is the  inhomogeneously transforming part of the connection 
under a  change of the coordinates.
 The contribution of  $B^\la_{\mu\nu}$ 
signifies   the ``hard''/kinetic extension to GR and 
clearly requires both  the local-Lorentz and quasi-affine tetrads, respectively,
$\vart^\al_\mu$ and  ${\zeta}{}^a_\mu$.

\subsubsection{Suppressed  contributions} 

At last,  the third  group of terms is as follows:  
\bea
\eta_{ab}&\to&{\zeta}_{\mu\nu} \equiv {\zeta}{}_\mu^a {\zeta}{}_\nu^b \eta_{ab}, \nn\\
\eta^{ab}&\to&{\zeta}^{\mu\nu} \equiv {\zeta}{}^\mu_a {\zeta}{}^\nu_b \eta^{ab}, \nn\\
  \det(\eta_{ab})&\to&  \zeta
  \equiv \det(\zeta_{\mu\nu})=\det({\zeta}{}^a_\mu)^2 \det(\eta_{ab}) ,\nn\\
H^a{}_b\equiv\theta^{ac}\eta_{cb}&\to&
H^\mu{}_\nu={\zeta}{}_a^\mu {\zeta}{}^b_\nu H^a{}_b =g^{\mu\la} \zeta_{\la\nu}.
\eea
Call $\zeta_{\mu\nu}$ the  quasi-Minkowskian metric. 
The  terms above  originate clearly from those in the affine realization 
space containing  $\eta_{ab}$ and imply thus 
an explicit affine symmetry violation.\footnote{In particular, this concerns
the quasi-Minkowskian measure  $\sqrt{-\zeta}$.}
Clearly,  the  tetrad ${\zeta}{}^a_\mu\equiv\partial_\mu Z{}^a$  defines  
the (patch-wise)  quasi-affine coordinates $z^a=Z{}^a(x)$,
wherein  $\zeta_{ab}(z)=\eta_{ab}$ and $\ga^c_{ab}(z)=0$. At 
that, the term $\ga^\la_{\mu\nu}$ eq.~(\ref{ga})  
defined originally through ${\zeta}{}^a_\mu$ (and its inverse) 
may equivalently  be presented as the Christoffel connection corresponding to the 
quasi-Minkowskian metric $\zeta_{\mu\nu}$, 
with $B^\la_{\mu\nu}$ getting explicitly the GC tensor. 
Nevertheless, 
the dependence on $\eta_{ab}$, in fact, drops off and the term  
$B^\la_{\mu\nu}$  does not violates  the affine symmetry.\footnote{A small
explicit violation of the affine symmetry through  $\eta_{ab}$ may, in principle,  serve
as a {\em raison d'\^{e}tre} inducing the affine SSB, the latter  surviving even after 
such a  ``seed''  violation dropping-off.} Finally remark that were $z^a=Z^a(x)$ 
not the dynamical but some prior/``absolute'' coordinates 
the affine-Goldstone NM would be either incomplete  (under retaining
only  the first  group of terms shown above, which are 
expressed entirely through the local-Lorentz tetrad~${\zeta}{}_\mu^\al$  
absorbing $Z{}^a$) or 
unclosed (under addition of the two last groups 
of terms dependent also  on   
the  quasi-affine tetrad~${\zeta}{}_\mu^a$).

\subsection{Second-derivative reduction}

The gravity action $I$ in a world manifold ${M^4}$ looks generically like
\be
I=\int_{M^4} L_G { \cal M} d^4 x,
\ee
with $L_G$ a GC scalar Lagrangian and  $ { \cal M} $  
a spacetime  measure,
i.e., a GC  scalar density of the proper weight 
for $I$ to be a GC scalar. 
 At that, due to a  prior freedom of redefining   the 
 Lagrangian there is an   arbitrariness of choosing the measure.
E.g., the latter may be defined  as $\sqrt{-g}$ 
or $\sqrt{-\zeta}$, or as  a combination of both.
By this token,  the action  for the pure gravity  
may be expressed without loss of generality
 through the 
metric $g_{\mu\nu}$ and the field  $H^\mu{}_\nu$   as follows:
 \be
I=\int L_G( \partial_\la g_{\mu\nu}, g_{\mu\nu}, H^\mu{}_\nu) \sqrt{-g}\, d^4 x.
\ee
The second-derivative  Lagrangian in the GC form now becomes  
\be
 L_G= L_g+ \ka_{\rm P}^2 K ( B^\la_{\mu\nu},  g_{\mu\nu},H^\mu{}_\nu)
-V(H^\mu{}_\nu),
\ee
with
\be
L_g= -\frac {1}{2}\ka^2_{\rm P}(1+ \ve_0(H)) R ,
\ee
where  $R$  is the conventional  Ricci scalar  constructed from the Christoffel
connection $\Ga^\la_{\mu\nu}(g_{\mu\nu})$.\footnote{By default, the indices 
of   $H^\mu{}_\nu=g^{\mu\la}\zeta_{\la\nu}$, $H^{-1}{}^\mu{}_\nu =
\zeta^{\mu\la}g_{\la\nu}  $, etc.,  
are raised or lowered, if necessary, 
through $g^{\mu\nu}$ or $g_{\mu\nu}$, respectively.}
Here we have  identified the affine SSB scale $\ka_0$ with the reduced Planck mass
$\kappa_{\rm P}=1/\sqrt{8\pi G_{\rm N}}$, where $G_{\rm N}$ 
is the Newton's constant.

The kinetic term $K$ looks  as  before 
\be\label{K'} 
K = \frac{1}{2} \sum_p  \ve_p(H)   {K}_p(B^\la_{\mu\nu}, g_{\mu\nu}),
\ee
with  the partial  contributions  as follows:
\bea
{K}_1= g ^{\mu\nu} B_{\mu\ka}^\ka B_{\nu\la}^\la,   && 
 {K}_2=g_{\mu\nu}  g^{\ka\la}g^{\rho\si}
B_{\ka\la}^\mu   B_{\rho\si}^\nu,\nn\\
 {K}_{3}=g^{\mu\nu} B_{\mu\nu}^\ka B_{\ka\la}^\la,      &&
  {K}_{4}=g_{\mu\nu}  g^{\ka\la}  g^{\rho\si} B_{\ka\rho}^\mu
B_{\la\si}^\nu    ,\nn\\  
 {K}_5=  g^{\mu\nu}  B_{\mu\ka}^\la  B_{\nu\la}^\ka.  && 
\eea
The potential  $V(H)$ remains as before, with $H^a{}_b$ 
substituted by $H^\mu{}_\nu=g^{\mu\la}\zeta_{\la\nu}$.
The simplest case is given under dependence of $K$ only on 
a scalar-graviton field
\be
\si  \equiv\ln (\det(H^\mu{}_\nu))^{-1/2}= \ln \sqrt{-g}/\sqrt{-\zeta},
\ee
with
$\ve_1\neq 0$ and  the rest of $\ve_p$'s being zero, so that
\be 
K= \frac{1}{2} \ve_1 (\si) K_1= 
\frac{1}{2} \ve_1 (\si) g^{\mu\nu} \partial_\mu\si\partial_\nu\si,
\ee
supplemented by a potential $V=V(\si)$.
Likewise, we could  reproduce  in the affine-Goldstone  
NM the more elaborate cases for the quartet-metric  gravity~\cite{Pir1, Pir2}.

\subsection{Weak-field limit}

Let now the dynamical gravity fields $Z^a$ and $g_{\mu\nu}$ are decomposed as 
\bea
Z{}^a&=&\bar Z^a+\zeta^a,\nn\\
g_{\mu\nu}&=&\bar g_{\mu\nu} + h_{\mu\nu},
\eea
with the backgrounds $\bar Z^a$  and $\bar g_{\mu\nu}$ 
supplemented by the small dynamical deviations $\zeta^a$ and  $ h_{\mu\nu}$,
respectively.
 Choosing the quasi-affine coordinates 
$z^a= \bar Z^a(x)$, so that $\bar \zeta_{ab}=\eta_{ab}$, and assuming for simplicity 
$\bar g_{ab}=\eta_{ab} $ we get   (operating  
indices now through $\eta_{ab}$ and  $\eta^{ab}$) 
in the linear approximation~(LA) as follows:
\bea
\zeta_{ab}&=&\eta_{ab}+\partial_a \zeta_b+\partial_b \zeta_a ,\nn\\
H_{ab}&=& \eta_{ab}-  h_{ab} +   \partial_a \zeta_b+\partial_b \zeta_a .
\eea
Likewise,  the kinetic term proves to exhibit  the similar  substitution: 
\be
h_{ab} \to   h'_{ab}  =  h_{ab} -   (\partial_a \zeta_b+\partial_b \zeta_a).
\ee
This means that  there takes place the Higgs-like  SSB of the quartet-metric  gravity, 
with  converting 
the originally  gauge components of 
the metric into  the physical ones to be treated as 
the  dark substances  of the Universe,  in addition to the tensor graviton.
Just to  illustrate a variety of the arising possibilities 
we consider in what follows the two  extreme cases
under choosing for simplicity the flat backgrounds: 
$\bar g_{ab}=\bar\zeta_{ab}=\bar H_{ab}=\eta_{ab}$.

\subsubsection{Pure-tensor gravity}

In one extreme case, imposing  on $\bar \ve_p\equiv  \ve_p(\bar H)$  the  constraints
\bea\label{e_t}
\bar \ve_1 =0,&&\bar\ve_2=-\bar\la,\nn\\
\bar \ve_3= \bar \ve_t-\bar\la,& &\bar \ve_4=\bar\la,\nn\\
\bar \ve_5=-\bar \ve_t+ 3\bar\la,& &
\eea
with $\bar \ve_t$ and $\bar\la$ some free parameters, 
we recover  for $h'_{ab}$ in LA the conventional 
GR Lagrangian  in an obvious  notation as follows:
\be
L_G= \frac{1}{8}\ka_{\rm P}^2 (1+\bar \ve_0 +\bar \ve_t)
\Big ((\partial_c h'{}^{ab})^2 -2(\partial_a h'{}^{ab})^2 
  +2\partial_a h'{}^{ab}  \partial_b h'{}^c_c  - (\partial_a h'{}^b_b)^2    \Big) ,
\ee
 without modification of the tensor gravity. At that, $\bar\la$ drops off completely, 
with $\bar\la=0$ in the simplest version.
To reproduce  GR in LA exactly we  should additionally put 
$\bar \ve_g\equiv \bar \ve_0 + \bar \ve_t=0$,
recovering  the Newton's gravity in LA precisely. 
The weak-field post-Newtonian contributions to $L_G$  impose, generally,  
some  restrictions on the 
left-out parameters $\bar\ve_t$ and $\bar\la$.
The deviations from the  relations shown above  would imply some additional  
kinetic contributions beyond GR already in LA, 
being, as it could be anticipated strongly suppressed. 
At last, the potential $V_t(h'^{ab})$  may be  
chosen so to recover in LA the Fiertz-Pauli term  
for the massive tensor graviton possessing, due to the Higgs-like mechanism,   
the descent massless limit. 
Nevertheless,  even at the zero deviations from GR in   LA
the full nonlinear theory  may still essentially deviate from GR 
 in the  strong-field limit producing some additional   restrictions/predictions.

\subsubsection{Scalar-tensor gravity}

In other  extreme case,  with  $\bar\ve_0=0$ and
\bea
\bar \ve_1 =\bar\ve_s,&&\bar\ve_2=-\bar\la,\nn\\
\bar \ve_3= -\bar\la,& &\bar \ve_4=\bar\la,\nn\\
\bar \ve_5= 3\bar\la,& &
\eea
where $\bar \ve_s$ is a free parameter 
corresponding to the addition of the scalar graviton $\si  =h'{}^c_c/2 \equiv h'/2 $ through 
\be
K=\frac {1}{8}\bar\ve_s \partial_a h' \partial^a h' 
\ee
without modifying tensor gravity.  With $\bar \la $  dropped=off completely, 
 one can put $\bar\la=0$  in the simplest version. 
This kinetic term should be supplemented by 
a potential $V_s(h')$ for the scalar-graviton mass and self-interactions.
{\em A priori}, there is also possible
to consider a  (putatively  problematic) vector graviton,
as well as a  mixture  of the different kinds of gravitons.
Such the additional gravity components,
having the NG origin common with the tensor graviton, 
may naturally  be treated as the gravitational dark substances 
(DM, DE, etc.) of the  Universe.

\section{Summary: NM vs.\ EFT  and beyond}

Let us briefly summarize the main advantages 
provided for  the quartet-modified  GR  
by  the  NM  vs.\  EFT frameworks:

\begin{itemize}
\item 
NM  justifies 
the extended set of the gravity fields, $g_{\mu\nu}$ and $Z^a$, 
as well as  the pattern of the basic symmetries: 
 GC and the global Poincare symmetry,  the latter 
introduced  in EFT {\em ad hock}.

\item 
NM refines the types of the gravity interactions 
and their hierarchy as  
originating from the enhanced, normal and suppressed terms  
in the affine realization space.

\item
NM ascribes the clear-cut physical meaning to the  Planck scale  $\ka_{\rm P}$, 
determining the strength of the gravity interactions, 
as a scale of  the affine SSB.

\item NM states the  unique tensor graviton  
and the 
putative  gravitational dark substances  (DM, DE, etc.)
as the emergent pNG bosons corresponding to the affine SSB.

\item
NM may ultimately serve as a  link  between the quartet-metric EFT of gravity 
below  the Planck scale  
and  the  producing NM underlying  theory above this scale, 
pointing, arguably,   towards the gravity and spacetime  
as  emerging  at~$\ka_{\rm P}$ due to the affine~SSB. 

\end{itemize}

Altogether, in the  paper there is  proposed  the 
route of the GR  modification, so to say, ``in width'' and  ``in depth''  
as follows: 
$$
\mbox{\dots}\stackrel{\ka_{\rm P}}{\longleftrightarrow}
\mbox{affine-Goldstone NM}
{\longleftrightarrow}
\mbox{(reduced) quartet-metric EFT}  
{\longleftrightarrow} \mbox{GR} . 
$$
For studying  the manifestations of  gravity 
it is sufficient to  stay  at the level of EFT,
but for a deeper revealing the nature  of gravity it is necessary
to adhere to the more profound  level given by  NM.
The respective two-level  phenomenon, being the affine-Goldstone in its  nature 
and the quartet-metric  in its appearance, may  be called   
the  affine-Goldstone/quartet-metric  gravity.
Further studying the latter, both theoretically and phenomenologically, 
to verify its validity  is urgent.

\paragraph{Acknowledgments} The author is  grateful to S.~S.\ Gershtein 
for  interest to the work and  encouragement, and  
to V.~V.\ Kabachenko for discussions.

\section*{Appendix: SSB  generalities}

\paragraph{Cosets} 

To describe  group-theoretically the SSB $G\to H$  of a global 
continuous  internal symmetry/group  $G$ to its subgroup  $H\subset G$ 
we should first clarify  the   notion  of   
the so-called  cosets.
A (say, left) coset of a subgroup $H$ in the group $G$ with respect to   
an  element $k\in G$ is defined generically as 
a subset (not, generally, a subgroup) of $G$  
 equivalent  to  the given $k$ {\em modulo}  the (right) multiplication 
 by any $h\in H$, i.e., $k\sim k h$.
The cosets are either identical or disjoint.
Each element of $G$ belongs to one and only one coset, 
with the cosets partitioning the group, i.e., 
the unification of all the cosets represents the whole group.
At that, the coset as a whole is uniquely determined  by any  its  
element chosen as a representative, say,  $k$.
The total set of the (left) coset representatives, $K=\{ k\}$, 
constitutes  the (left) coset space $K=G/H$.
The coset space  ultimately serves as a space for NRs and NMs to implement SSB,
with the cosets providing  a  ``language'' for SSB.

\paragraph{SSB and (pseudo-)Nambu-Goldstone bosons} 

Now, if $G$ is a symmetry group of a physical system, 
let $| I\!\! >$ be the system  ground state (``vacuum'') 
invariant under $G$, i.e., $g | I\!\! > =  | I\!\! >$ for any
$g\in G$. Let now there takes place SSB $G\to H$ meaning that 
the invariance of the vacuum  lowers up to $H\subset G$, i.e., 
only $h | I\!\! > =  | I\!\! >$ for any $h\in H$. 
In this  case,  for an arbitrary $g\in G$, with 
the decomposition $g=kh$, $k\in K$ and $h\in H$, 
there takes place $g | I\!\! > =  k | I\!\! >\equiv  | k \!\! >\neq   | I\!\! >$.
This corresponds to   SSB in  the NG mode, with the appearance 
of a set of the degenerate vacua   $ | k \!\! >$, $k\in K$, 
the  excitation  of which corresponds to the  physical NG bosons.
Under an explicit violation of $G$, the NG bosons become, in fact, 
the pseudo-NG (pNG) ones. In these terms, 
the  field variable on a flat spacetime   
describing the  SSB $G\to H$ in the  NG mode may conventionally
 be chosen as the  coset (local) representative $k(x)\in K$ for   $x \in R^4$.
 
\paragraph{Nonlinear realizations}

Further, the result of the action of a group element $g\in G$ on a (left) coset 
representative $k\in K$, $g k$,  being a group element 
may as well  be decomposed as $g k=k' h$,
with a new $k'\in K$ and some $h=h(g,k )\in H$.
This implies that  the coset representative  
$k$ transforms under $G$  nonlinearly as 
$$
g\ : \ k \to  k'=        g kh^{-1}(g, k).
$$
For $g=h_0\in H$, there  can be shown that $h(h_0,k)= h_0$ and, in particular,
$h(I,k)=I$, with $k'=k$.
This defines  NR of the group $G$ on the (left) coset space $K=G/H$
for  the SSB $G\to H$.

\paragraph{Nonlinear models} 

And finally, a NM describing  the NG field  $k\in K=G/H$  
appearing due to  SSB $G\to H$ 
for a pair of the global internal groups $H\subset G$ 
is a  specific field theory for the coset (local)  representatives 
$k(x)\in K$, $x\in R^4$.  Such a theory may be defined   
by an action $S=\int  L (k, \psi)d^4 x$,
with a Lagrangian $L$  invariant under 
the nonlinearly realized/``hidden'' symmetry $G$.
At that, a generic matter field $\psi$ transforms as  
a linear representation  $\rho$  of the residual (classification) subgroup $H\subset G$  
through  $h(g,k)$ determined above:
$$
g \ : \psi\to \psi'=\rho(h(g,k))\psi,
$$
with the NG field  $k$ being thus omnipresent 
(what, by the way,  is characteristic of graviton).
Under $g=h_0\in H$, this becomes the conventional 
linear representation  $\rho(h_0)$ of $H\subset G$.
This generically defines  NM for  the SSB $G\to H$ built on NR of the group $G$, 
reducing to the  linear representation when restricted by the unbroken subgroup $H\subset G$. 

The  affine-Goldstone NM  for gravity
considered  in the given paper  corresponds
(with a modification)
to the symmetry pattern $G/H=GL(4,R)/SO(1,3)$.   
As the pNG bosons there serve  here  the (in a  general case, massive) tensor graviton 
and the additional  gravity  components, to be associated  ultimately 
with  the gravitational DM and DE.


\end{document}